\documentstyle[10pt,aaspp4,tighten,flushrt]{article}

\newcommand{\kms}{\mbox{${\rm\,km\,s}^{-1}$}}
\newcommand{\kpc}{\mbox{$\rm\,kpc$}}
\newcommand{\mpc}{\mbox{$\rm\,Mpc$}}
\newcommand{\msun}{\mbox{$M_\odot$}}
\newcommand{\myr}{\mbox{$\rm\, Myr$}}
\newcommand{\beq}{\begin{equation}}
\newcommand{\eeq}{\end{equation}}

\lefthead{Statler}
\righthead{NGC 3379}

\begin{document}

\title{The Shape and Orientation of NGC 3379: Implications for Nuclear
Decoupling}

\author{Thomas S. Statler}
\affil{Department of Physics and Astronomy, Ohio University, Athens, OH 45701,
USA; tss@coma.phy.ohiou.edu}

\vskip -1.9in {\hfill \sl Astronomical Journal, Jan.\ 2001, in press}
\vskip 1.8in

\begin{abstract}

The intrinsic shape and orientation of the elliptical galaxy
NGC 3379 are estimated by dynamical modeling. The maximal ignorance shape
estimate, an average over the parameter space, is axisymmetric
and oblate in the inner parts, with an outward triaxiality gradient.
The $1\sigma$ limits on total-mass triaxiality $T$ are $T<0.13$ at
$0.33\kpc$ and $T=0.08\pm 0.07$ at $3.5\kpc$ from the center. The luminous
short-to-long axis ratio $c_L=0.79 {+0.05 \atop -0.1}$ inside $0.82\kpc$,
flattening to $c_L=0.66 {+0.07 \atop -0.08}$ at $1.9\kpc$. The results
are similar if the galaxy is assumed to rotate about its short axis.
Estimates for $c_L$ are robust, but those for $T$ are dependent on whether
the internal rotation field is disklike or spheroid-like. Short-axis
inclinations $i$ between $30\arcdeg$ and $50\arcdeg$ are preferred for
nearly axisymmetric models; but triaxial models in high inclination are
also allowed, which can affect central black hole mass estimates.
The available constraints on orientation rule out the possibility that the
nuclear dust ring at $R\approx 1\farcs 5$ is in a stable equilibrium in
one of the galaxy's principal planes. The ring is thus a {\em decoupled nuclear
component\/} not linked to the main body of the galaxy. It may be
connected with ionized gas that extends to larger radii, since the
projected gas rotation axis is near the minor axis of the ring. The gas
and dust may both be part of a strongly warped disk; however, if caused
by differential precession, the warp will wind up on itself in a few
$10^{7}$ years. The decoupling with the stellar component suggests
that the gas has an external origin, but no obvious source is present.

\end{abstract}


\section{Introduction}

Like many astronomers, the ``standard elliptical galaxy'' NGC 3379 appears
outwardly normal but, deep inside, has serious issues. To the casual observer
this object is as ordinary as can be. Its optical surface brightness
distribution follows an $r^{1/4}$ law over a span of 10 magnitudes
(de Vaucouleurs \& Capaccioli \markcite{dVC79}1979). Its isophotes are
almost perfectly elliptical and aligned (Peletier et
al.\ \markcite{Pel90}1990). Its optical colors are typical of an
old population, and color gradients are small (Goudfrooij
et al.\ \markcite{Gou94}1994). It rotates slowly about its apparent minor axis
(Davies \& Illingworth \markcite{DaI83}1983, Davies \& Birkinshaw
\markcite{DB88}1988). It shows no sign of current or past interactions
(Schweizer \& Seitzer \markcite{ScS92}1992), and contains only a tiny
amount of detectable interstellar matter, seen only in H$\alpha +$[\ion{N}{2}]
emission (Macchetto et al.\ \markcite{Mac96}1996) and dust absorption
(van Dokkum \& Franx \markcite{vDF95}1995, Michard \markcite{Mic98}1998).

On closer examination, the photometric and kinematic structure of the
galaxy are puzzling. Deviations of the surface brightness
profile from the $r^{1/4}$ law, though small, are significant
(Capaccioli et al.\ \markcite{CaH87}1987). Based on photometric similarities
to NGC 3115, Capaccioli et al.\ \markcite{CVHL}(1991) propose that NGC
3379 is a highly triaxial S0 galaxy seen at low ($31\arcdeg$) inclination. 
The orientation and shape of the galaxy are difficult to constrain, however.
Van der Marel et al.\ \markcite{vdM90}(1990), using stellar
kinematic data and explicitly axisymmetric models, infer an inclination
of $60\arcdeg$, at which the galaxy would be intrinsically almost as
round as it appears. Statler \markcite{Sta94}(1994c, hereafter S94)
considers triaxial
models, introducing a methodology that relies on multi-position-angle
observations of the mean velocity field. These models rule out the axis ratios
advocated by Capaccioli et al.\ \markcite{CVHL}(1991) at $98\%$
confidence,\footnote{The model of Capaccioli et al.\ is not internally
consistent due to unfortunate trigonometric errors (S94, \S\
5.2).}\ but allow a wide variety of flattened nearly axisymmetric, or
rounder triaxial shapes, and a correspondingly wide variety of
inclinations.

Detailed kinematic studies of both stars (Bender, Saglia, \& Gerhard
\markcite{BSG94}1994; Statler \& Smecker-Hane \markcite{SS99}1999,
hereafter SS99) and planetary nebulae (Ciardullo, Jacoby, \& Dejonghe
\markcite{CJD93}1993) show a richer kinematic structure than one might
have expected from older data. Inflections in the rotation and
dispersion profiles are suggestive of a two-component structure, and are
similar to those of S0 galaxies (Fisher \markcite{Fis97}1997). SS99
suggest a low ($30\arcdeg$) inclination and a {\em weakly\/} triaxial
S0-like shape on the basis of these similarities. At large radii the
stellar dispersion profile joins with that of the planetary nebulae,
which is consistent with a {\em constant\/} mass-to-light ratio out to
$r \approx 10\kpc$ (Ciardullo, Jacoby, \& Dejonghe \markcite{CJD93}1993).

\begin{figure}[t]{\epsfxsize=3in\hfil\epsfbox{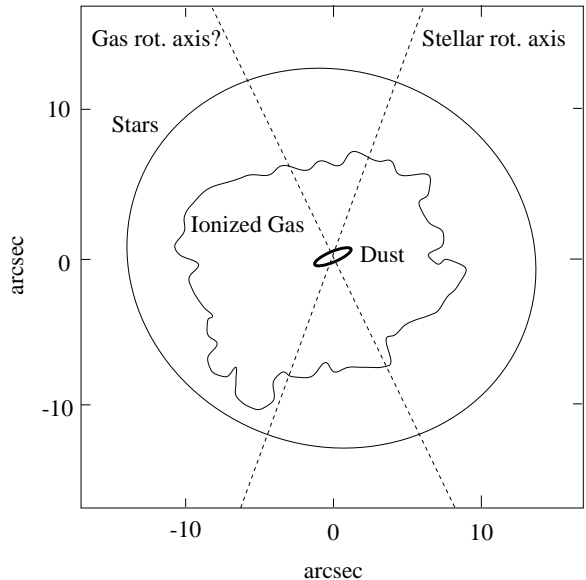}}
\caption{\footnotesize
Sketch of the central regions of NGC 3379, oriented with North at the top and
East at left. A typical stellar isophote is shown
by the big ellipse. The apparent rotation axis of the stars coincides very
closely with the morphological minor axis, shown as the thick dotted line.
The convoluted outline shows the outermost H$\alpha$+\ion{N}{2} isophote
plotted by Machetto et al.\ (1996). The gas rotation is measured only
within $3\arcsec$ of the center, and the rotation axis is poorly determined.
Gas kinematics are consistent with the rotation axis lying along the apparent
minor axis (light dotted line) of the central dust ring (thick ellipse). Both
the stellar and gas velocity fields are such that the right side of the figure
is at positive velocities and the left side at negative.
\label{f.sketch}}
\end{figure}

More recently, the nuclear regions of NGC 3379 have caught the attention
of a number of workers, particularly as a likely home for a supermassive
black hole. Kinematic data from the ground and from {\em HST\/} imply
a central mass in the vicinity of $10^8 \msun$ (Magorrian et
al.\ \markcite{Mag98}1998; Gebhardt et al.\ \markcite{Geb00a}2000a,
hereafter G00), placing it roughly in the middle of the
$M_{\rm BH}$-$L_{\rm bulge}$ (Magorrian et al.\ \markcite{Mag98}1998) and
$M_{\rm BH}$-$\sigma_{\rm bulge}$ (Ferrarese \& Merritt
\markcite{FeM00}2000, Gebhardt et al.\ \markcite{Geb00b}2000b)
correlations, but the black hole mass is dependent, to within factors of
order unity, on the orientation of the galaxy.
The nuclear dust ring (\markcite{Geb00a}G00, Pastoriza et al.\
\markcite{Pas00}2000, hereafter P00) seen on arcsecond
scales and the slightly more extended H$\alpha +$[\ion{N}{2}] emission
(Macchetto et al.\ \markcite{Mac96}1996, \markcite{Pas00}P00) might be useful in
constraining the orientation, if it could be assumed that the
interstellar material were in an equilibrium configuration in one of the
galaxy's principal planes. However, there seems not to be compelling
justification for such an assumption, since the apparent principal axes
of the dust and ionized gas are aligned neither with the stellar
principal axes nor with each other. Figure \ref{f.sketch} shows a sketch of
these components and their relation to the starlight.

In this paper, I revisit the problem of constraining the
shape and orientation of the main body of NGC 3379, using the newest
stellar kinematic data (\markcite{SS99}SS99, Gebhardt \& Richstone
\markcite{GeR00}2000). The approach follows that of \markcite{Sta94}S94,
as modified by Statler, Dejonghe, \& Smecker-Hane \markcite{SDS99}(1999,
hereafter SDS). The data are of sufficiently high quality to attempt to
recover not just an average shape, but the shape {\em profile\/} over a
full decade in radius. The presentation is arranged as follows: Section
2 describes the photometric and kinematic data used in the modeling.
Section 3 gives an overview of the method, and details where the present
treatment differs from previous applications. Sections 4 and
5 describe the results for the estimates of the intrinsic shape profile
and the orientation, respectively. Section 6 discusses the implications
of the results for the structure and evolution of NGC 3379's nuclear
regions, and Section 7 restates the major conclusions.

\section{Observational Data\label{s.data}}

The handling of the observational material parallels that
in earlier work. Kinematic data consist of mean velocities
measured along several long-slit cuts through the galaxy center.
Data for P.A.'s 205, 250, 295, and 340 come from \markcite{SS99}SS99;
their published
velocities for $6\arcsec \leq R \leq 78\arcsec$ are reflected about the
center and then averaged over pairs of adjacent bins, giving 9 annular zones
of mean radius $6\farcs 6$, $9\farcs 0$, $11\farcs 4$, $13\farcs 8$,
$16\farcs 2$, $19\farcs 8$, $26\farcs 4$, $38\farcs 4$, and $69\farcs 0$.
Data inside $6\arcsec$ are not used because the models perform poorly in
regions where the rotation curve is steeply rising.
Additional data for P.A.'s 180, 225, and 235 are taken from Gebhardt \&
Richstone \markcite{GeR00}(2000)
and linearly interpolated onto the same radial scale.

Ellipticities and major axis position angles are drawn from the $R$ band
photometry of Peletier et al.\ \markcite{Pel90}(1990)
and averaged over the same radial
bins used for the kinematic data, weighted by the inverse square of the
published errors. The adopted errors in the rebinned data represent the
larger of the mean observational error per data point and the standard
deviation in the bin.

The complete set of observational data used in the modeling is summarized in
Table 1.

\begin{figure}[p]{\epsfxsize=3in\hfil\epsfbox{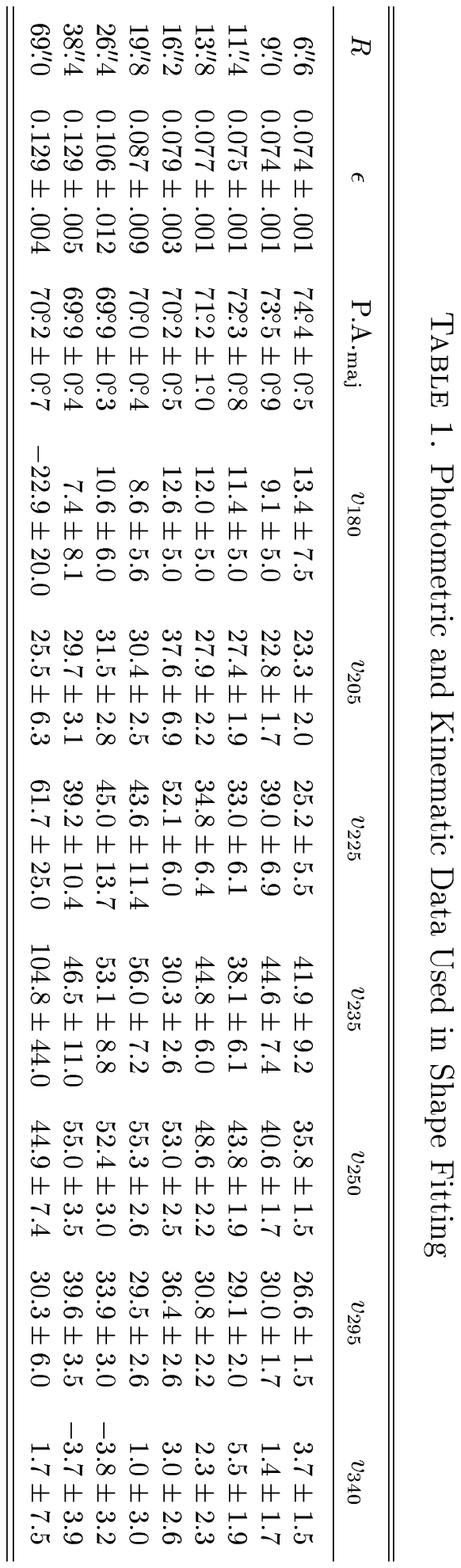}}
\end{figure}

\section{Models and Methods\label{s.method}}

\subsection{Thumbnail Sketch}

The shape-fitting method and the models on which it is based have been
described exhaustively in earlier papers (Statler \markcite{PapI}1994a,
Statler \& Fry \markcite{PapII}1994, Statler \markcite{PapIII}1994b, hereafter
Papers I, II, and III; \markcite{S94}S94, \markcite{SDS99}SDS).
The method finds, in a probabilistic sense,
the boundaries between the major circulating orbit families using the
2-dimensional radial velocity field and its
alignment with the surface brightness distribution. The locations of
these boundaries and the apparent shape of the galaxy then constrain the
underlying 3-dimensional mass distribution. The analysis of the velocity
field proceeds by comparison with dynamical models, which are
constructed by solving the equation of continuity for the mean stellar
velocities using confocal streamlines linked to the triaxiality $T$ of
the total mass distribution. The triaxiality is defined in terms of the axis
ratios of the equidensity surfaces by $T\equiv (1-b^2)/(1-c^2)$, where $b$
and $c$ are the middle-to-long and short-to-long axis ratios, respectively;
corresponding axis ratios for the luminosity distribution are labeled by
$b_L$ and $c_L$. The major assumption of the models is that rotation of
the galaxy arises from internal streaming in a non-rotating potential.
Certain other simplifying assumptions are made to facilitate the
projection along the line of sight. The models include a number of
adjustable parameters, which allow a wide array of dynamical configurations
to be computed efficiently. These parameters are described in Paper
I, and collectively abbreviated by $\mbox{\boldmath $d$}$.

The statistical estimate of the shape or orientation of the galaxy is
obtained by Bayesian methods. Computing a large grid of models
results in a multidimensional likelihood function
$L(T,c_L,\Omega,\mbox{\boldmath $d$})$ in the space of shape $(T,c_L)$,
orientation $\Omega$, and dynamical parameters $\mbox{\boldmath $d$}$.
The likelihood is multiplied by
a prior, or model parent distribution, which is isotropic in
orientation, but which may reflect assumptions about the dynamics. The
product is integrated over all unwanted parameters to give a posterior
probability distribution for the parameters of interest. A shape estimate
is thus a probability density $P(T,c_L)$. As in earlier work it is assumed
that the isodensity surfaces of mass and luminosity have the same shape
at the same radii; tests in Paper III showed that the distribution in
$(T,c_L)$ is still robust when this assumption is mildly violated.

The 9 annular zones of the galaxy are initially modeled independently,
then combined assuming that the galaxy's principal axes are intrinsically
aligned throughout (see Appendix A of \markcite{SDS99}SDS).
If it is explicitly assumed that the
triaxialities of the luminous and total mass distributions are the same, then
the shape profile can also be required to reproduce the observed isophotal
twist (which is small in this case). Except where specifically noted,
all of the results in this paper include the isophotal twist constraint,
but are not significantly altered by ignoring it.

\subsection{Modifications to the Basic Procedure}

\subsubsection{Model Grid}

A preliminary set of models shows that, as in earlier work, the
results for NGC 3379 are very insensitive to the parameters $k$ and $l$,
which describe the radial dependence of luminosity density and mean
rotation assumed in the projection integrals (see Paper I, equation [32]).
Rather than use a grid of values for each parameter as in previous
cases, I simply fix the values at $k=3$, $l=0$ for the final modeling.

\begin{figure}[t]{\epsfxsize=4.5in\hfil\epsfbox{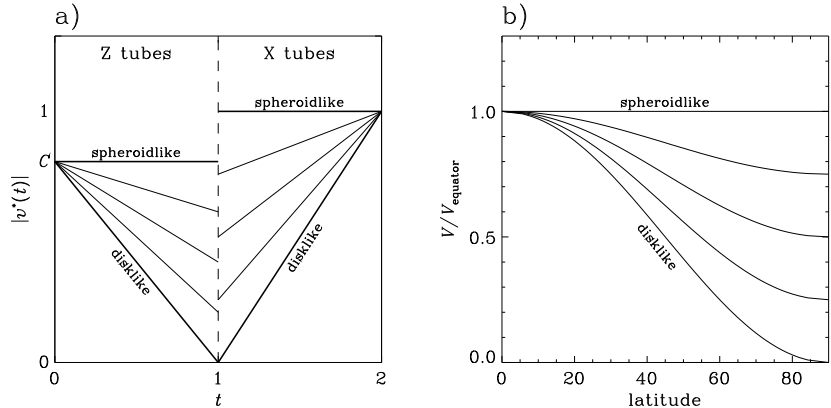}}
\caption{\footnotesize
($a$) Adopted forms for the $v^\ast(t)$ function, which gives the mean
internal velocity crossing the $x$-$z$ plane as a function of a scaled latitude
$t$. The $x$ and $z$ axes correspond to $t=0$ and $t=2$, respectively;
$t=1$ marks the boundary between short-axis and long-axis tubes. {\em
Heavy lines\/} show the extreme disklike and spheroid-like cases used
in earlier work; {\em light lines\/} show the intermediate cases added
for the present models. ($b$) Translation of $v^\ast(t)$ for short-axis
tubes in an axisymmetric oblate model into true mean velocity as a
function of latitude above the equatorial plane.
\label {f.vstar}}
\end{figure}

The same preliminary models show a complicated dependence on the function
$v^\ast(t)$, which describes the latitude dependence of the mean
streaming velocity across the $x$-$z$ plane.\footnote{The $x$, $y$,
and $z$ axes correspond to the long, middle, and short axes of the
model.}\ The variable $t$ is a scaled latitude, defined (equation
[9] of Paper II) so that $t=0$ and $t=2$ correspond to the $x$ and $z$
axes, respectively, and the boundary between short-axis and long-axis
tube orbits\footnote{For brevity I also use the synonyms ``Z tubes''
and ``X tubes.''}\ is at $t=1$. In previous papers, $v^\ast(t)$ in each
interval was assumed to be either constant or
linearly decreasing to zero at $t=1$, as shown by the heavy lines in
Figure \ref{f.vstar}a; these limiting cases are referred to as
``spheroid-like'' and ``disklike'' rotation. For NGC 3379, I find it
necessary to include the intermediate cases. Again, I take $v^\ast(t)$
to be linear in each interval, with maxima at $t=0$ and $t=2$, decreasing
to $(\case{1}{4},\case{1}{2},\case{3}{4})$ of the maximum at $t=1$;
these functions are shown as the light lines in Figure \ref{f.vstar}a.
To aid intuition, Figure \ref{f.vstar}b shows how these choices translate
into the actual latitude dependence of the velocity, for streaming about
the short axis in an oblate axisymmetric model. Note that in a triaxial
model the character of the long and short axis streaming can be completely
independent.

\subsubsection{Penalty for Steep Shape Gradients\label{s.gradpenalty}}

Combining the separate annular fits as prescribed in \markcite{SDS99}SDS
effectively
merges the 9 annuli as if they were separate galaxies
seen from the same direction. This prescription ignores the physical
constraint that the axis ratios of the isodensity surfaces should change
continuously---and, for {\em bona fide\/} ellipticals, slowly---with $r$.
To remedy this I introduce a set of penalty factors of the form
\beq
\exp\left[-{(T_{i+1}-T_i)^2+(c_{L,i+1}-c_{L,i})^2 \over 2 \sigma_g^2}\right]
\eeq
for each pair of adjacent annuli ($i=1,\ldots,8$), with the constant
$\sigma_g=0.118$. This choice implies that a complete change of
shape (i.e., oblate to prolate, or sphere to disk) between the
innermost and outermost annuli would be penalized as {\em a priori\/}
unlikely at the $4\sigma$ level.

\subsubsection{Parent Shape Distribution}

In previous papers a flat prior distribution in $(T,c_L)$ was
adopted for modeling individual systems. However, a realistic estimate
of the parent shape distribution has now been presented by Bak \&
Statler \markcite{BS00}(2000, hereafter BS), derived from
the 13-galaxy Davies \& Birkinshaw \markcite{DB88}(1988)
sample, which includes NGC
3379. BS show that their ``maximal ignorance'' distribution, obtained from an
unweighted sum over all parameters $\mbox{\boldmath $d$}$, gives an
ellipticity distribution completely consistent with the 165-galaxy
compilation of Ryden \markcite{Ryd92}(1992).
For this work I adopt the \markcite{BS00}BS maximal
ignorance result (their Figure 2a) as the prior shape distribution.

\newpage
\begin{figure}[ht]{\epsfxsize=5.0in\hfil\epsfbox{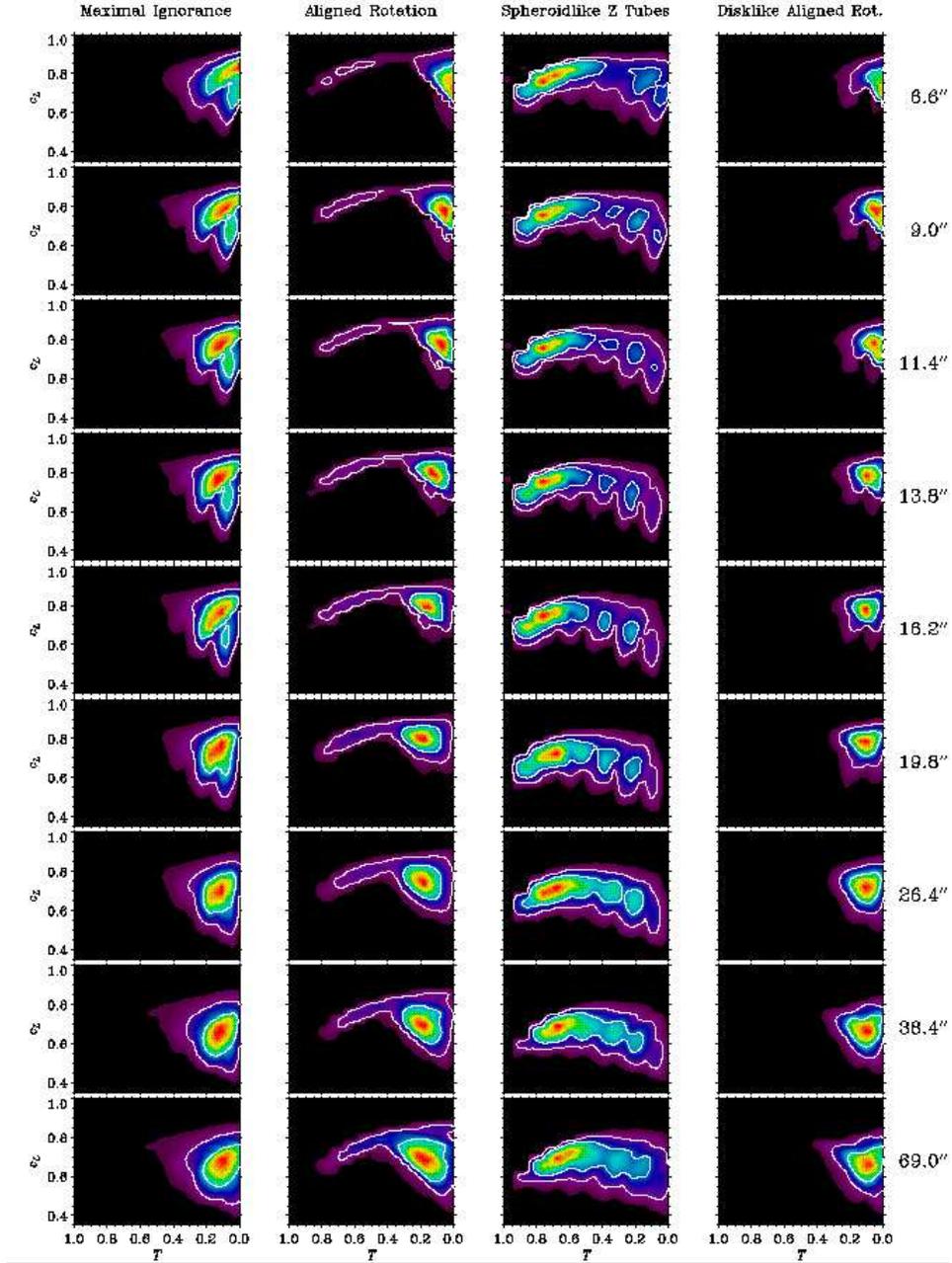}}
\caption{\footnotesize
Shape profile estimates for NGC 3379. Each column shows an estimate
using a different dynamical prior; each small panel shows the Bayesian
probability in total mass triaxiality $T$ and luminous axis ratio $c_L$
at one radius (labeled at far right). Contours mark the $68\%$ and $95\%$
highest posterior density (HPD) regions. {\em First column\/}: maximal
ignorance estimate, an unweighted average over the dynamical parameter
space. {\em Second column\/}: estimate using only models that rotate
about their short axes. {\em Third column\/}: estimate with spheroid-like
rotation in Z tubes (see \S\ \protect{\ref{s.shape}} of text). {\em
Fourth column\/}: estimate using only models with disklike rotation
about their short axes.
\label {f.shapes}}
\end{figure}

\newpage
\section{Intrinsic Shape\label{s.shape}}

Formally, the shape profile for the galaxy is a joint probability
distribution in 18 variables. To make this more
manageable, I follow \markcite{SDS99}SDS and project the distribution into the
$(T,c_L)$ plane for each annulus. The resulting 9 marginal distributions
are estimates of the mean shape in each annulus,
using the data from all radii but independent of the actual shapes of the
other annuli. Figure \ref{f.shapes} shows four representative results.
Each small panel shows the $(T,c_L)$ plane, with oblate spheroids at the
right edge, prolate spheroids at the left edge, and spheres at the
top. The marginal posterior density $P(T,c_L)$ is displayed on a linear
color scale, with contours enclosing the 68\% and 95\% highest posterior
density (HPD) regions. Each column shows one complete shape profile,
with the mean radii of the 9 annuli indicated along the far right margin.
The columns differ in the assumed prior distribution in the dynamical
parameters $\mbox{\boldmath $d$}$.

The small ellipticity and the symmetric, well-aligned velocity field
of NGC 3379 make it, ironically, a difficult object to model. Broadly
speaking, there are three types of models that fit the data (though
there are no clear divisions and so this trichotomy is artificial):
(1) fairly round models with small but nonzero triaxiality, seen at moderate
inclination;
(2) somewhat flatter nearly oblate models, seen at lower inclination; and
(3) highly triaxial models with lines of sight in the $x$-$z$ plane.
The first group of models is the most successful, and dominates the total
probability in an unweighted sum over all models. This sum gives the maximal
ignorance estimate, shown in the first column of Figure \ref{f.shapes}.
In this estimate, the galaxy is most probably axisymmetric at small
radii, and shows a slight increase in triaxiality and flattening toward
larger $R$. The contribution from the second group of models can be seen as
the weaker second peak near $T\approx 0.07$ for $R<20\arcsec$, and the third
group is responsible for the tenuous tail extending toward $T=0.5$.
The second column of Figure \ref{f.shapes} shows the shape estimate
under the assumption of zero intrinsic kinematic misalignment,
i.e., rotation purely about the short axis. This
result is obtained by omitting models with net rotation in the
X tube orbits. The most probable shape at each radius agrees fairly
closely with that in the maximal ignorance result, with a slightly
larger outward triaxiality gradient. The tail toward large $T$ is
more prominent, but contains a small fraction of the total
probability. Both this and the maximal ignorance result imply that NGC
3379 is weakly triaxial and probably close to oblate.

The third group of models becomes particularly problematic
in some parts of the $\mbox{\boldmath $d$}$ parameter space. For lines
of sight near the $x$-$z$ plane, the apparent kinematic misalignment is
always small, and one can freely vary the axis ratios while preserving the
projected shape, without moving the apparent rotation axis. A model of
almost any triaxiality can be made to fit the data in such an orientation by
fine-tuning the $v^\ast(t)$ function (Fig.\ \ref{f.vstar}).
As a result, restricting the dynamical prior to certain regions of
parameter space can give very different answers for the inferred triaxiality.
Two extreme cases are shown in the third and fourth columns of Figure
\ref{f.shapes}. In the third column, the models are restricted to those
in which the rotation in Z tubes is moderately
spheroid-like, corresponding to the second line from the top in the left
half of Figure \ref{f.vstar}. No constraint is put on the X tubes. The
result is dominated by prolate-triaxial models viewed in the $x$-$z$ plane.
In the fourth column, the aligned-rotation models from the second
column are restricted only to those with disklike rotation. The
disappearance of the distribution's large-$T$ tail confirms that
spheroid-like rotation makes the highly triaxial models viable.

\begin{figure}[t]{\epsfxsize=4in\hfil\epsfbox{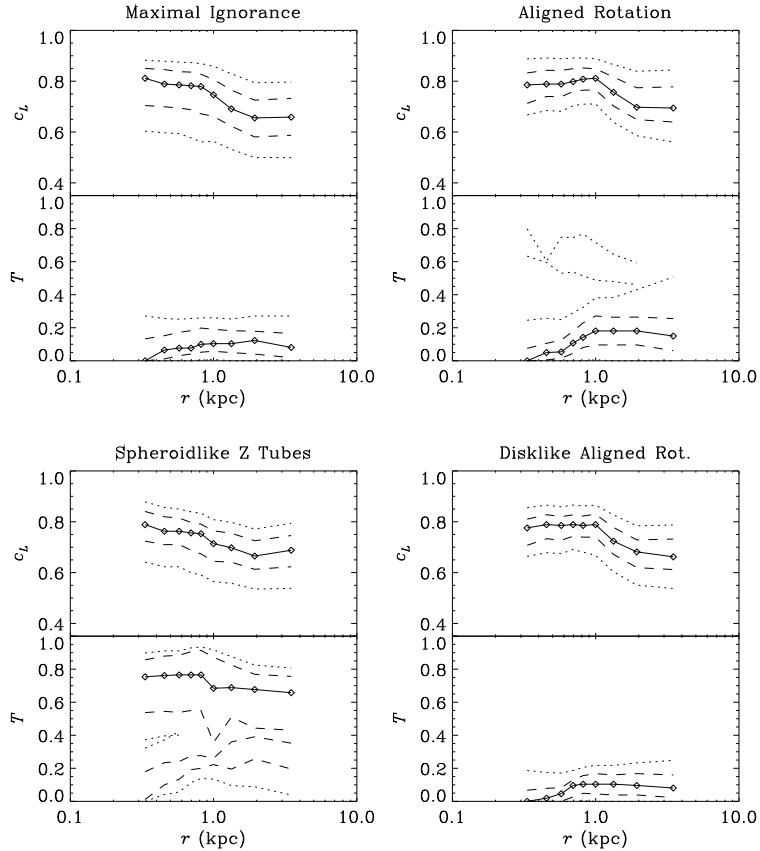}}
\caption{\footnotesize
Radial profiles of triaxiality and flattening for the same four cases
shown in Fig.\ \protect{\ref{f.shapes}}. Profiles are determined from
1-dimensional marginal distributions, hence these are not joint
$(T,c_L)$ profiles. {\em Points and solid lines\/},
maximum-posterior-probability values; {\em dashed lines\/}, crossings
of the $68\%$ HPD level ($1\sigma$ error region); {\em dotted lines\/},
crossings of the $95\%$ HPD level ($2\sigma$ error region).
\label {f.tcprofiles}}
\end{figure}

To help in understanding the radial shape profiles, I project the
distributions in Figure \ref{f.shapes} onto the $T$ and $c_L$ axes and
plot the 1-dimensional marginal distributions against $r$.
Figure \ref{f.tcprofiles} shows the profiles of $T$ and $c_L$ for the
same four cases shown in Figure \ref{f.shapes}.
The plotted points connected by solid lines show the
most-probable values (i.e., peaks of the marginal distribution), and the
dashed and dotted lines indicate crossings of the 68\% and 95\% HPD
levels. The radial scale has been set using a distance of $10.4\mpc$
(Ajhar \clearpage

\noindent
et al.\ \markcite{Ajh97}1997). One can easily see the tendency for
outwardly-increasing triaxiality in 3 out of 4 cases, as well as the
effect of altering the dynamical prior. Even though the maximal
ignorance result gives a formally narrow error region, the systematic
uncertainties associated with disklike or spheroid-like
rotation are large in comparison; in other words, the triaxiality
profile is {\em prior dominated\/}, and should not be considered a
robust result. On the other hand, the radial run of the short-to-long
axis ratio $c_L$ {\em is\/} robust, and not sensitive to the dynamical
prior. The maximal ignorance profile is flat or slowly declining for
$r\lesssim1\kpc$, then changes over a factor $\sim 2$ in radius
from the inner value of $c_L \approx 0.8$ to an outer value
$c_L\approx 0.7$. This transition is driven primarily by the photometric
ellipticity gradient, though modulated by kinematics, and occurs more
abruptly in the models that rotate purely about their short axes.

\section{Orientation\label{s.orientation}}

A different projection through the parameter space can give a probability
distribution for the orientation of NGC 3379. The central points of
interest here are the inclination of the (probably axisymmetric) inner
region and the relation of the stars to the nuclear dust ring (Figure
\ref{f.sketch}). The ring has a semi-major axis
of approximately $1\farcs 5$, with the minor axis along P.A. $206\pm 5$,
roughly $45\arcdeg$ away from the minor axis of the starlight.
If the ring is circular, its aspect ratio indicates that its symmetry axis
is inclined to the line of sight by $73\arcdeg\pm 5\arcdeg$.

\begin{figure}[t]{\epsfxsize=4in\hfil\epsfbox{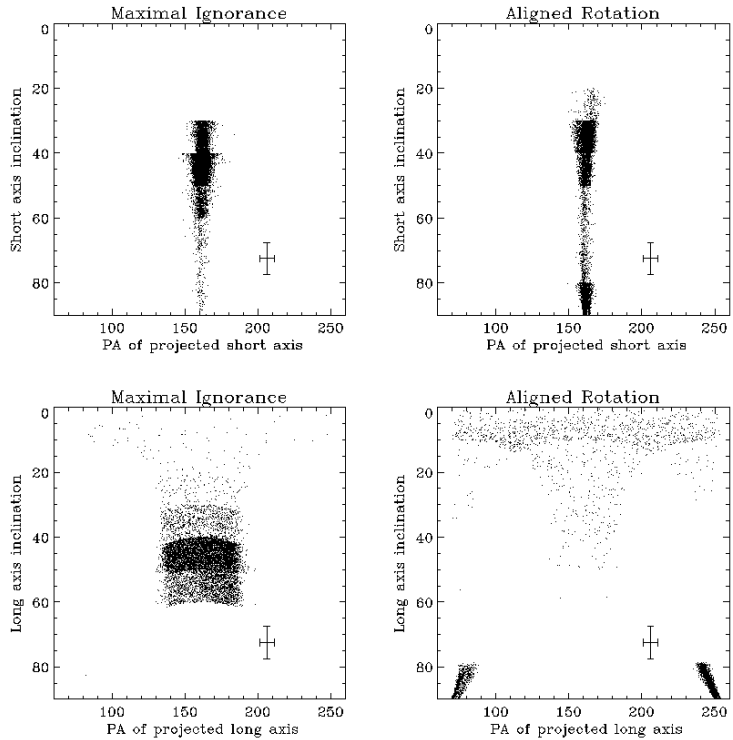}}\\
\caption{\footnotesize
{\em Top row\/}: Joint distribution in short axis inclination
to the line of sight and projected short axis position angle, for ({\em
left\/}) maximal ignorance and ({\em right\/}) rotation about the short
axis. {\em Error bars\/} mark the orientation required if the nuclear
dust ring is assumed to lie in the plane normal to the short axis.
Horizontal banding reflects the coarse ($10\arcdeg$) angular grid in the
polar angle $\theta$.
{\em Bottom row\/}: Same, but for the orientation of the long axis;
{\em error bars\/} mark the orientation required if the
dust ring is assumed normal to the long axis.
\label {f.axisdist}}
\end{figure}

In a non-tumbling triaxial potential, a massless ring can be stable in the 
planes normal to the long or short axes.\footnote{A tumbling rate
sufficient to create a significantly tilted equilibrium for the nuclear ring
would place corotation well within the main body of the galaxy, so this
case is excluded.}\ The latter case would seem the
more likely, if the ring dynamics are coupled to the stellar dynamics.
This would require P.A.\ 206 to correspond to the direction of
the projected short axis, and a $73\arcdeg$ inclination of this axis to
the line of sight. In the top two panels of Figure \ref{f.axisdist} I
show joint probability distributions in projected short axis P.A.\ and
short axis inclination calculated from the models, for the maximal ignorance
and aligned rotation cases. A rejection method has been used to populate
the particle-density plots.
The inclination is not well constrained, but the P.A.\ of the
projected short axis lies within $\pm 10\arcdeg$ of the apparent minor axis, at
better than $3\sigma$ confidence. The error bars show the orientation
required if the dust ring were normal to the short axis; it is clear that 
this configuration is not viable.

One might be concerned that this result could be altered
if the luminous and total-mass triaxialities were different, in which
case constraining the mass triaxiality profile to reproduce the observed
isophotal twist would be inappropriate. Removing this constraint widens
confidence intervals on the projected short axis P.A.\ by
about 50\%, but the basic result is unchanged.

It is worth reiterating that the short-axis inclination is not well
constrained by the data, even though the flattening is fairly well determined.
This can be a counterintuitive result, if one's intuition is based on
axisymmetric models. Even a weakly triaxial model does not project to a given
apparent shape at a unique inclination. Since $T$ is poorly
constrained, the inclination is as well, and suffers from the
same systematic uncertainties. Modelers are cautioned not to think of
the inclination as being set by the short-to-long axis ratio.
This is a valid inference only if one asserts via an {\em ad hoc\/}
prior that the galaxy is axisymmetric.

If the dust ring were normal to the long axis, then this axis would have to
fall at P.A.\ 206 in projection and lie at the right inclination. The bottom
row of Figure \ref{f.axisdist} shows the results from the models. The
constraints are much weaker than for the short axis, and also more
dependent on the dynamical prior. However, the distributions shown are
typical in that different subsets of the $\mbox{\boldmath $d$}$
parameter space do not populate regions of the figure that are
unpopulated in these examples. The same is true for models in which the
isophotal twist constraint is lifted. The orientation required by the dust lane
falls squarely in an unpopulated region; I conclude that
a configuration with the ring about the long axis is also ruled out.

The last equilibrium refuge for the dust lies in making the galaxy exactly
axisymmetric in the inner parts and placing the dust in a polar ring at
the correct azimuth, where it would be neutrally stable. From the {\em
HST\/} photometry of \markcite{Geb00a}G00, the
ellipticity just outside the ring is $0.11$, and the minor axis P.A.\ is 
$161\arcdeg$, which must equal the P.A. of the projected short axis. To
match the observed ring orientation and aspect ratio requires the short
axis to be inclined $24\arcdeg$ to the line of sight. At this inclination
all oblate objects---even razor-thin disks---project to ellipticities
$<0.11$. On the other hand, the required inclination is very sensitive
to the ring parameters; pushing both the ring P.A. and inclination by
$1\sigma$ in the favorable directions changes the galaxy's inclination
to $33\arcdeg$, at which a flattening of $c_L=0.54$ will project to the
correct ellipticity. Pushing by $1.5\sigma$ in both directions allows
$c_L=0.66$ to fit the inner photometry at an inclination of $38\arcdeg$,
which is well within the viable region for oblate models found in Section
\ref{s.shape}. A polar ring configuration is therefore not ruled out.

\section{Discussion\label{s.discussion}}

By fitting dynamical models to the stellar kinematics and photometry,
I have shown that NGC 3379 is most probably axisymmetric and oblate in
its inner parts, with a weak but significant outward triaxiality
gradient.\footnote{
It is worth noting that this gradient is required both by the observed
isophotal twist and by the twist of the velocity field, which are in
opposite directions (see Fig.\ 6a of SS99). Relaxing the
requirement that the models match the isophotal twist gives the same 
$T$ gradient as that in the top row of Fig.\ \ref{f.tcprofiles}, but at lower
significance.}\ There are, however, regions of parameter
space where the best models are prolate-triaxial. Thus the estimate of
small triaxiality is dependent on the dynamical prior, and, pending a
better understanding of the internal velocity fields of ellipticals,
should not be considered robust.

Nonetheless, this result bolsters the use of oblate axisymmetric models
for constraining the mass of the putative central black hole
(Magorrian et al.\ \markcite{Mag98}1998, \markcite{Geb00a}G00).
Even though the present models use data only
from $R > 6\arcsec$, there is no sign of serious photometric
(\markcite{Geb00a}G00) or
kinematic (\markcite{SS99}SS99) twists interior to this radius that
could signal a loss of axisymmetry. If axisymmetry is asserted {\em a
priori\,}, and the axis ratios assumed constant from $R=6\arcsec$
inward, the maximal-ignorance flattening $c_L=0.81 {+.05 \atop -.1}$
would imply an inclination of $40 {+8 \atop -4}\arcdeg$. At this
inclination, the models of \markcite{Geb00a}G00
imply a central dark mass of $2\times
10^8 \msun$. At the same time, one must
be conscious of the highly triaxial solutions, particularly those at
high inclinations, which would imply smaller central masses than
the less-inclined axisymmetric models. G00's
best-fit black hole mass drops by a factor of 2 as they tilt their
oblate models from $50\arcdeg$ to $90\arcdeg$ (their preferred
inclination). In addition, triaxial
systems observed down their long axes can appear atypically ``hot'' because of
velocity anisotropy, requiring less dark mass for a given
observed dispersion if triaxiality persists to small radii.

The present results for the shape of NGC 3379 are consistent with those
found earlier using older kinematic data (\markcite{Sta94}S94,
Fig.\ 8; \markcite{BS00}BS,
Fig.\ 4). On the other hand, they do not strongly favor
the argument of \markcite{SS99}SS99 that the galaxy is a
weakly triaxial, slightly
puffy S0-like system seen at low inclination. SS99 make this argument based 
on photometric and kinematic similarities with NGC 3115 (Capaccioli et
al.\ \markcite{CVHL}1991, Fisher \markcite{Fis97}1997). In the present
models, configurations that are highly flattened ($c_L<0.5$) and nearly
oblate are viable and do contribute to the total probability density, but
only at a low level. Moreover, their contribution comes almost entirely
from the most spheroidlike models, whereas one would expect more
disklike rotation if the galaxy were a two component disk-bulge
system. Alternatively, if
one {\em believes\/} that the nuclear dust is in a polar ring, this
configuration requires an inclination $< 43\arcdeg$ ($3\sigma$
confidence), consistent with that argued for by \markcite{SS99}SS99.

But there is no compelling reason for the dust ring to be polar. The
evolution of such an object is not like the evolution of the large polar
rings commonly observed (e.g., Whitmore et al.\ \markcite{Whi90}1990)
and modeled (e.g. Sackett et al.\ \markcite{Sac94}1994). A nuclear ring
is immersed in the rotating stellar population, and will be torqued
by dynamical friction. One should expect a nuclear polar ring to
secularly tilt out of the meridional plane and begin to differentially precess
in the oblate potential. Big polar rings can be stabilized by self gravity
(Sparke \markcite{Spa86}1986, Arnaboldi \& Sparke \markcite{ArS94}1994),
but this is not an option for the NGC 3379 ring, which
obscures only $\sim 1\%$ of the galaxy's $V$-band surface brightness.
Assuming a Galactic dust-to-gas ratio, this $A_V$ implies a mass only
of order $100\msun$, compared with the $\sim 10^8\msun$ enclosed by the ring.

Regardless of whether the nuclear ring is strictly polar or merely
misaligned with the symmetry axes, the results of Section
\ref{s.orientation} show clearly that it is effectively decoupled from
the main body of the galaxy. This echoes the original finding of Davies
\& Birkinshaw \markcite{DB88}(1988) of a lack of correlation between
the radio and optical axes of radio ellipticals. In general, recent
imaging studies (van Dokkum \& Franx \markcite{vDF95}1995,
Carollo et al. \markcite{Car97}1997) have
found a bewildering variety of nuclear dust morphologies in ellipticals,
which seem totally uncorrelated with properties of the host galaxies.
For NGC 3379, the observed dust surely cannot have come directly from the
stellar population, whose mass distribution and angular momentum are extremely
well ordered and not aligned with the dust ring. For one of the most
ordinary and quiescent ellipticals in the sky to contain a misaligned---and,
presumably, rapidly evolving---non-stellar component is remarkable.

The nature and history of the dust ring may be related to the ionized
gas that extends from the center out to $R\approx 8\arcsec$ (Figure
\ref{f.sketch}; Macchetto
et al. \markcite{Mac96}1996). \markcite{Pas00}P00 obtain
H$\alpha +$[\ion{N}{2}] rotation curves for $R\lesssim 3\arcsec$ along
P.A.\ 70, 115, and 160. The maximum observed velocity is approximately
$220\kms$ on P.A.\ 70 and 115, and roughly half this on P.A. 160, which
is the stellar minor axis. These amplitudes are consistent with the gas
rotation axis lying parallel to the axis of the dust ring.
From the very round H$\alpha
+$[\ion{N}{2}] isophotes and an assumption that the gas distribution is
circular, \markcite{Pas00}P00
infer an inclination of $25\arcdeg$ for the ionized disk.
However, this inclination implies gas rotation velocities in excess of
$500\kms$, which seems unrealistic considering the $\sim 200\kms$
dispersion of the stars.

More likely, the ionized gas lies in a strongly warped disk, in which
case the outer H$\alpha +$[\ion{N}{2}] isophotes may not reflect the disk's
inclination at the smaller radii where \markcite{Pas00}P00
measure the rotation. 
The dust ring could then be part of the same structure, either marking
the inner edge of the disk or merely made prominent by its low inclination.
Macchetto et al.\ estimate the ionized disk to be only a few $10^3 \msun$
in mass. It is far from clear whether a low mass, low density warped disk
could be a long-lived object. In a slightly oblate isothermal system with
velocity dispersion $\sigma$, inclined, nearly circular orbits of radius
$r$ precess at a frequency given in the epicycle approximation by
\beq
\omega_p = {\sqrt{2} \sigma \over r} \left({1\over q} -1 \right),
\eeq
where $q$ is the flattening of the potential (Statler et
al.\ \markcite{SSC96}1996). The preferred flattening $c_L=0.8$ gives
$q=0.93$ if mass follows light, from which the precession period is
\beq
{2\pi\over\omega_p} \approx 14 \left({R \over 1\arcsec}\right) \myr.
\eeq
To warp an initially coplanar disk by $90\arcdeg$ between $R=1\arcsec$ and
$R=5\arcsec$ by differential precession thus requires only about $4\myr$. 
Significantly later than this, say, by $20\myr$, the disk will have
evolved into a multiply-wrapped sheet, which may be inconsistent
with the observed velocity field. This simple calculation ignores
the effect of dynamical friction against the stars, which will
simultaneously try to align the disk's angular momentum with the stellar
rotation axis.

\section{Conclusions\label{s.summary}}

Dynamical models have been fit to the stellar kinematics and surface
photometry of NGC 3379, in order to constrain the intrinsic shape
profile and orientation of the main body of the galaxy. The Bayesian
``maximal ignorance'' shape estimate, which reflects an unweighted
average over the dynamical parameter space, indicates that the galaxy is
most probably axisymmetric and oblate in its inner parts, with a slight
but significant outward triaxiality gradient. The formal $1\sigma$
limits on triaxiality $T$ are $T<0.13$ at $R=6\farcs 6$ ($0.33 \kpc$) and
$T=0.08\pm 0.07$ at $R=69\arcsec$ ($3.5\kpc$). The most probable
short-to-long axis ratio of the luminosity distribution is
$c_L=0.79 {+0.05 \atop -0.1}$ for $R<16\arcsec$ ($0.82\kpc$), flattening
beyond this point and reaching $c_L=0.66 {+0.07 \atop -0.08}$ at
$R=38\arcsec$ ($1.9\kpc$). The results are similar, with slightly
steeper gradients in $T$ and $c_L$, if the galaxy is assumed {\it a
priori\/} to rotate about its short axis. While the estimates for $c_L$
are robust, those for $T$ are dependent on the degree to which the
internal rotation field is disklike or spheroid-like. There are regions
in the dynamical parameter space where the preferred shapes are
prolate-triaxial. Thus the triaxiality estimate should not be considered
robust until the dynamical prior can be better constrained on physical
grounds.

The inclination of the short axis to the line of sight is not well
determined, because the triaxiality is not well determined.
Inclinations between $30\arcdeg$ and $50\arcdeg$ are preferred. However,
highly triaxial models in high inclination are also allowed, which can
affect central black hole mass estimates (Magorrian et al.
\markcite{Mag98}1998, \markcite{Geb00a}G00) by factors of a few. The
position angle of the projected short axis on the sky {\it is\/} well
constrained, and lies within $10\arcdeg$ of the photometric minor axis at
better than $3\sigma$ confidence. Similar, though weaker, constraints
can be placed on the P.A.\ and inclination of the long axis. These
constraints rule out the possibility that the
nuclear dust ring at $R\approx 1\farcs 5$ lies in one of the planes
normal to either axis. Stable equilibria for planar gas disks 
in triaxial potentials exist only in these planes. Alternatively, if
$T\equiv 0$ and the short axis inclination is $<43\arcdeg$ ($3\sigma$
limit), the dust could be in a neutrally stable equilibrium in a polar
ring. However, even in this configuration the ring would be torqued by
dynamical friction against the rotating stellar population, which would
act toward aligning the ring with the stellar rotation axis. The ring 
is clearly a {\em decoupled nuclear component\/} whose orientation is
not linked to the main body of the galaxy.

The dust ring may be connected with the ionized gas that extends out to
$R\approx 8\arcsec$ (Macchetto et al. 1996, \markcite{Pas00}P00).
The projected gas rotation axis is near the minor
axis of the ring, though the H$\alpha +$[\ion{N}{2}] isophotes are very
round. It is possible that the gas and dust are both part of a strongly
warped disk. However, the evolution and expected lifetime of such a
structure are problematic. If the warp is caused by differential
precession, one should expect the entire structure to wind up on itself
in $\sim10^{7}$ years. The decoupling with the stellar component would
suggest that the gas has an external origin. But no trace of recent
interaction or accretion is seen at larger radii, where it would
presumably be much longer lived. At very large radii ($\gtrsim
10\arcmin$), NGC 3379 and NGC 3384 are collectively surrounded by a ring of
\ion{H}{1} (Schneider \markcite{Sch85}1985) whose connection with these
and neighboring galaxies is far from clear (Schneider
\markcite{Sch89}1989). Whether the intergalactic ring could be a source
of material for occasional accretion events, and how this material might
evolve dynamically as it accreted into the nuclear regions, are topics
for future work.

\acknowledgments

Many thanks to Karl Gebhardt for providing data in advance of publication
and for helpful comments. An anonymous referee also kindly suggested several
improvements to the paper.
This work was supported by NSF CAREER grant AST-9703036.

\end{document}